\documentstyle[12pt]{article}

\begin{document}
\mbox{ }
\rightline{UCT-TP-252/98}\\
\rightline{October 1998}\\
\vspace{3.5cm}
\begin{center}
{\Large \bf Chiral Sum Rules and Duality in QCD}\footnote{Work supported
in part by the Volkswagen Foundation}\\
\vspace{.5cm}
{\bf C. A. Dominguez$^{(a)}$, and K. Schilcher$^{(b)}$}\\[.5cm]
$^{(a)}$Institute of Theoretical Physics and Astrophysics\\
University of Cape Town, Rondebosch 7700, South Africa\\[.5cm]
$^{(b)}$Institut f\"{u}r Physik, Johannes Gutenberg-Universit\"{a}t\\
Staudingerweg 7, D-55099 Mainz, Germany
\end{center}
\vspace{.5cm}
\begin{abstract}
\noindent
The ALEPH data on the vector and axial-vector spectral functions, extracted
from tau-lepton decays, is used in order to test local and global duality,
as well as a set of four QCD chiral sum rules. These are the 
Das-Mathur-Okubo sum rule, the first and second Weinberg sum rules, and a 
relation for the electromagnetic pion mass difference. We find  these 
sum rules to be poorly saturated, even when the upper limit in the 
dispersion 
integrals is as high as $3 \; \mbox{GeV}^{2}$. Since perturbative QCD, plus
condensates, is expected to be valid for $|q^{2}| \geq \cal{O}$$(1 \;
\mbox{GeV}^{2})$ in the whole complex energy plane, except in the vicinity 
of the right hand cut, we propose a modified set of sum rules with weight 
factors that vanish at the end of the integration range on the real axis.
These sum rules are found to be precociously saturated by the data to a
remarkable extent. As a byproduct, we extract for the low energy 
renormalization constant $\bar{L} _{10}$ the value $- 4 \bar{L} _{10}
= 2.43 \times 10^{-2}$, to be compared with the standard value
$- 4 \bar{L} _{10} = (2.73 \pm 0.12) \times 10^{-2}$. This in turn leads
to a pion polarizability $\alpha_{E} = 3.7 \times 10^{-4} \;\mbox{fm}^{3}$. 
\end{abstract}
\newpage
\setlength{\baselineskip}{1.5\baselineskip}
\noindent
There is a set of sum rules involving the difference between vector and 
axial-vector spectral functions, first discovered in the framework 
of current algebra \cite{CA}, which are now understood as consequences 
of the underlying chiral symmetry of QCD. These spectral functions are
related to the discontinuities in the complex energy plane of the
two-point functions involving the vector and axial-vector currents
\begin{eqnarray*}
 \Pi_{\mu \nu}^{VV} (q^2) = i \; \int \; d^4 \; x \; e^{i q x} \; \;
  <0|T(V_{\mu}(x) \; \; V_{\nu}^{\dagger}(0))|0> \; 
\end{eqnarray*}
\begin{equation}
  = \; (- g_{\mu \nu} \; q^{2} + q_{\mu} q_{\nu}) \; \Pi_{V} (q^{2}) \; ,
\end{equation}
\begin{eqnarray*}
 \Pi_{\mu \nu}^{AA} (q^2) = i \; \int \; d^4 \; x \; e^{i q x} \; \;
  <0|T(A_{\mu}(x) \; \; A_{\nu}^{\dagger} (0) )|0> \; 
\end{eqnarray*}
\begin{equation}
  = \; (- g_{\mu \nu} \; q^{2} + q_{\mu} q_{\nu}) \; \Pi_{A} (q^{2})
  - q_{\mu}  q_{\nu} \; \Pi_{0} (q^{2})  \; ,
\end{equation}
where $V_{\mu}(x) = :\bar{q}(x)  \gamma_{\mu}  q(x):$,
$A_{\mu}(x) = :\bar{q}(x)  \gamma_{\mu}  \gamma_{5} q(x):$, and
$q=(u,d)$.  In this note we will study specifically the chiral correlator
$\Pi_{V-A} \equiv \Pi_{V}-\Pi_{A}$. This correlator vanishes identically in 
the chiral limit ($m_{q}=0$), to all orders in QCD perturbation theory.
Renormalon ambiguities are thus avoided. Non-perturbative contributions
to this two-point function start at dimension $d=6$, and involve the
four-quark condensate. In the factorization approximation, and at high
momentum transfer, this contribution is given by
\begin{equation}
\Pi_{V-A} (q^{2})  \; 
= \; \frac{32 \pi}{9} \;
\frac{\alpha_{s} <\bar{q} q>^{2}}{q^{6}} \;
\left\{ 1 +  \frac{\alpha_{s} (q^{2})}{4 \pi} \;
\left[ \frac{247}{12} + {\rm ln} \left( \frac{\mu^{2}}{-q^{2}} \right)
\right] \right\} \; + {\cal O} (1/q^{8}) \;.
\end{equation}
The low energy behaviour of $\Pi_{V-A} (q^{2})$ is governed by chiral 
perturbation theory; to one loop order this is given by
\begin{equation}
\Pi_{V-A} (q^{2})  \; = \frac{f_{\pi}^{2}}{q^{2}-\mu_{\pi}^{2}}\;
+ \; \frac{1}{48 \pi^{2}} \left[ \sigma^{2} \left( \sigma \ln \frac{\sigma 
- 1} {\sigma + 1}  + 2 \right) + \frac{2}{3} \right] + 4 \bar{L}_{10} \; ,
\end{equation}
where $\mu_{\pi}$ is the pion mass, $f_{\pi}$ the pion decay constant,
$f_{\pi} = 92.4 \pm 0.26 \; \mbox{MeV}$ \cite{PDG},
$\sigma \equiv (1 - 4 \mu_{\pi}^{2}/q^{2})^{1/2}$, and $\bar{L}_{10}$
is the scale independent part of the coupling constant of the relevant 
operator in the
$\cal{O}$$(E^{4})$ Lagrangian of chiral perturbation theory \cite{GL}.
The latter is related to the following combination of hadronic parameters
\begin{equation}
\bar{L}_{10} = - \frac{1}{4} \left[ \frac{1}{3} f_{\pi}^{2} <r_{\pi}^{2}>
-F_{A} \right] \; ,
\end{equation}
where $<r^{2}_{\pi}>$ is the electromagnetic mean squared radius of the pion,
$<r^{2}_{\pi}> = 0.439 \pm 0.008 \; \mbox{fm}^{2}$ \cite{AMEN},
and $F_{A}$ is the axial-vector coupling measured in radiative pion decay,
$F_{A} = 0.0058 \pm 0.0008$ \cite{PDG}. In the sixties, a number of sum
rules were derived for the discontinuity of $\Pi_{V-A} (q^{2})$, in
the complex energy plane, from current algebra and an assumed asymptotic
behaviour. 
The first sum rule is that of Das-Mathur-Okubo (DMO) \cite{DMO}
\begin{equation}
W_{0} \;  \equiv \; \int_{0}^{\infty} \; \frac{ds}{s} \;
\left[ \rho_{V} (s) - \rho_{A} (s) \right]
   = \; \frac{1}{3} \; f_{\pi}^{2} \; <r_{\pi}^{2}> - F_{A} = - 4 \; 
   \bar{L}_{10} = (2.73 \pm 0.12) \times 10^{-2}\; ,
\end{equation}
where $\rho_{V,A}(s)$ are related to
the imaginary parts of $\Pi_{V,A}$ , and normalized according to
\begin{equation}
\rho_{V,A} (s) = \frac{1}{8 \pi^{2}} \; \left[ 1 + {\cal{O}} \; (\alpha_{s}) 
\right] \; .
\end{equation}
The second and third relations we 
consider are the  first and second Weinberg sum rules \cite{WSR}
\begin{equation}
W_{1} \equiv
\int_{0}^{\infty} \; ds \;
\left[ \rho_{V} (s) - \rho_{A} (s) \right] = f_{\pi}^{2} \; ,
\end{equation}
\begin{equation}
W_{2} \equiv
\int_{0}^{\infty} \; ds \; s \;
\left[ \rho_{V} (s) - \rho_{A} (s) \right] = 0 \; .
\end{equation}
The final sum rule is a relation for the electromagnetic pion mass 
difference \cite{PMD}
\begin{equation}
W_{3} \equiv
\int_{0}^{\infty} \; ds \; s \;{\rm ln} \;
\left( \frac{s}{\mu^{2}} \right) \;
\left[ \rho_{V} (s) - \rho_{A} (s) \right]
= - \frac{16 \; \pi^{2} \; f_{\pi}^{2}}{3 \; e^{2}} \;
\left( \mu_{\pi^{\pm}}^{2} \; - \mu_{\pi^{0}}^{2} \right) \; . 
\end{equation}
Notice that this sume rule is actually independent of the arbitrary
scale $\mu$, by virtue of the second Weinberg sum rule, Eq.(9).
The above set of sum rules are exact relations in QCD, in
the chiral $SU(2)_{L} \times SU(2)_{R}$ limit, i.e. for vanishing up- 
and down-quark masses \cite{RA}.\\

With the advent of the ARGUS data on semi-leptonic tau-lepton decays
\cite{ARGUS}, a reconstruction of the vector and axial-vector spectral
functions became possible, albeit in the restricted kinematical range
limited by the tau-lepton mass, thus allowing for the above sum rules
to be confronted with experiment. This was first attempted in \cite{PS}.
Shortly after, an improved parametrization of the tau-lepton decay data
was discussed in \cite{CADS1}, and used in \cite{CADS2} to check the first
three sum rules above. Since kinematically $s < M_{\tau}^{2} < \infty $,
the question is how well does one actually verify these sum rules. It
turns out that by invoking quark-hadron duality, all three sum rules do
effectively become Finite Energy Sum Rules (FESR). In other words,
the upper limit of integration may be taken at a finite energy squared
$s = s_{0}$, where the asymptotic freedom thershold $s_{0}$ signals
the end of the resonance region and the start of the hadronic continuum,
which is well approximated by perturbative QCD. The conclusions of
\cite{PS} and \cite{CADS2}  were that the ARGUS data appeared to
saturate the sum rules reasonably well. In this paper we
make use of the recent, and much more accurate, ALEPH data \cite{ALEPH} 
on the vector and axial-vector spectral functions, extracted from
tau-lepton decays, to test the validity of (a) local and global
duality, and (b) the four chiral sum rules Eqs.(6),(8)-(10)(for
earlier related work see e.g. \cite{CONOS}).\\

The concept of duality is well defined in QCD. The starting point is
Wilson's Operator Product Expansion (OPE), extended beyond perturbative 
QCD by the addition of power corrections involving quark and gluon vacuum
condensates \cite{SVZ}. This OPE is expected to be valid in the whole
complex energy plane, except in the resonance  region.
In the space-like region ($q^{2} < 0$) the relevant scale of the OPE is
determined by $\Lambda_{QCD} \simeq 300 - 400$ MeV, on the perturbative
side, and by the scale of the condensates, also roughly 300 MeV, on the
non-perturbative side. It is therefore reasonable to use the OPE in the
region $- q^{2} \geq 1 \; \mbox{GeV}^{2}$. However, in the time-like region 
the relevant scale is much higher, as it is set by the masses of resonances
in the range $1 - 2 \; \mbox{GeV}$. In this region, the use of the OPE may
be further impaired by instanton and threshold effects. Hence, the OPE
may be used in the time-like region only at much higher scales than in
the space-like region. To account for this fact, one distinguishes different
types of duality, to wit. (i) Local duality
\begin{equation}
{\rm Im} \; \Pi \; (s) \; \simeq \; {\rm Im} \; \Pi \; (s)|_{\rm QCD} \; ,
\end{equation}
valid  in QCD at very high $- q^{2}$. (ii) Global Duality, which is an
integral statement based on Cauchy's theorem
\begin{eqnarray*}
\int_{s_{\rm th}}^{s_{0}} \; \frac{1}{\pi} \;
f (s) \; {\rm Im} \; \Pi (s) \; ds \; = \; - \frac{1}{2 \pi i} \;
\oint \; f (s) \; \Pi (s) \; ds
\end{eqnarray*}
\begin{equation}        
   \simeq \; - \; \frac{1}{2 \pi i} \;
\displaystyle{\oint_{|s| = s_{0}}} \; f (s) \; \Pi (s)|_{\rm QCD} \; ds \; .\\ 
\end{equation}
Alternatively, since Cauchy's theorem is valid in perturbative QCD,
\begin{equation}
\int_{s_{\rm th}}^{s_{0}} \; \frac{1}{\pi} \;
f (s) \; {\rm Im} \; \Pi (s) \; ds \; = \;  \int_{s_{\rm QCD}}^{s_{0}} \; 
\frac{1}{\pi} \; f (s) \; {\rm Im} \; \Pi (s)|_{\rm QCD} \; ds \; .
\end{equation}
In the above equations, $s_{th}$ is the physical threshold,
$s_{QCD}$ the QCD threshold, and $s_{0} > s_{th}$ is the so-called
asymptotic freedom threshold. The latter must be chosen large enough so
that $\Pi (s) \simeq \Pi (s)_{QCD}$ on the circle $|s| = s_{0}$.
The two equations above hold only if the weight function f(s)
is analytic in the domain $|s| \leq s_{0}$; for instance, if f(s) is a 
polynomial.
Global duality, therefore, relates weighted integrals of spectral functions
(connected to cross sections through the optical theorem) to integrals
of the QCD correlator over a large circle. Global duality in the interval
$s_{th} - s_{0}$ implies that local duality should be valid for $s > s_{0}$.
This may be seen by invoking Cauchy's theorem for a contour made of two
circles of radii $s_{0}$ and $s_{1} > s_{0}$, together with the connecting
lines along the cut. In practice, global duality may often hold to a much 
better degree than local duality, because oscillations can lead to
cancellations. (iii) Finally, we wish to introduce a form of global duality,
let us call it {\it restricted global duality}, expected to set in much
sooner than ordinary global duality. This can be accomplished if the weight
functions in Eq.(12) are constrained to vanish at $s = s_{0}$. A well
known example is the total hadronic decay rate of the $\tau$-lepton.
Below, we shall make use of this form of duality to write four modified
chiral sum rules, as alternatives to Eqs. (6), (8)-(10).\\
A few comments on the above chiral sum rules are in order. The left
hand side of Equations (8) and (9) can be related in a straightforward
manner to contour integrals of the QCD correlator. The same is true
of Eq.(6), provided one takes into account that $\Pi(0)$ is known, i.e.
\begin{equation}
\Pi_{V-A}(0) \; = - \; \frac{f_{\pi}^{2}}{\mu_{\pi}^{2}} 
 + 4 \; \bar{L}_{10} \; .
\end{equation}
In principle, duality arguments cannot be applied to Eq.(10) because of
the presence of the logarithm. As this is only mildly singular, and it
can be approximated rather well in the integration interval by an
analytic function, one could still invoke duality. However, one should
keep in mind that in this case duality may only be  approximately valid.
From a theoretical point of  view, the four chiral sum rules are expected 
to be saturated for relatively low values of $s_{0}$,
on account of the very rapid fall-off  of the correlator, cf. Eq.(3). \\

In Fig. 1 we show the ALEPH experimental data points \cite{ALEPH} on the
vector and axial-vector spectral functions, together with our fit
to these data (solid lines) and the perturbative QCD prediction
(dash line). The normalization of these
spectral functions has been changed (Eq.(7) has been multiplied by
$4 \pi^{2}$) in order to simplify comparisons with Ref. \cite{ALEPH}.
Experimental errors are small, of the size of the
dots in the figure, except near the end point. Even after taking these
errors into account, the agreement between data and theory, which measures
the validity of local duality, is not quite satisfactory. Furthermore,
because of cancellations, the errors on the difference $\rho_{V} - \rho_{A}$ 
should be smaller than the vector and axial-vector spectral function 
errors taken in quadrature. In view of this,
one would expect a similar unsatisfactory saturation of the four chiral
sum rules, Eqs.(6), (8)-(10), and  in fact, this is what we find. Before
illustrating our results, let us write down a set of four chiral
sum rules modified according to the principle of restricted global
duality introduced above
\begin{equation}
\bar{W}_{0} \;  \equiv \; \int_{0}^{\infty} \; \frac{ds}{s} \;
(1 - \frac{s}{s_{0}})\;
\left[ \rho_{V} (s) - \rho_{A} (s) \right]
   =- 4 \;\bar{L}_{10} \;- \; \frac{f_{\pi}^{2}}{s_{0}}
    \; ,
\end{equation}
\begin{equation}
\bar{W}_{1} \equiv
\int_{0}^{\infty} \; ds \;  (1- \frac{s}{s_{0}})\;
\left[ \rho_{V} (s) - \rho_{A} (s) \right] 
= f_{\pi}^{2} \; ,
\end{equation}
\begin{equation}
\bar{W}_{2} \equiv
\int_{0}^{\infty} \; \frac{ds}{s} \; (1- \frac{s}{s_{0}})^{2}\;
\left[ \rho_{V} (s) - \rho_{A} (s) \right] \;  = \;
- 4 \bar{L}_{10} - 2 \frac{f_{\pi}^{2}}{s_{0}} \; ,
\end{equation}
\begin{equation}
\bar{W}_{3} \equiv
\int_{0}^{\infty} \; ds \; s \;{\rm ln} \;
\left( \frac{s}{s_{0}} \right) \;
\left[ \rho_{V} (s) - \rho_{A} (s) \right]
= - \frac{16 \; \pi^{2} \; f_{\pi}^{2}}{3 \; e^{2}} \;
\left( \mu_{\pi^{\pm}}^{2} \; - \mu_{\pi^{0}}^{2} \right) \; ,
\end{equation}
where Eq.(15) is a combination of the DMO sum rule and the first 
Weinberg sum rule, Eq.(17) is a combination of the DMO sum rule and
the first and second Weinberg sum rules,
and the arbitrary scale in Eq.(18) has been fixed
to $s_{0}$, which becomes the upper limit of integration in the four sum
rules. We now show that these modified sum rules are 
saturated far better than the original sum rules Eqs.(6), (8)-(10).
In Fig.2 we plot the left hand side (l.h.s.) of
Eq.(6) computed using the fit to the data (curve(a)), and the right hand
side (r.h.s.) (curve(b)). Agreement with the data can be considerably
improved by rescaling the r.h.s. of Eq.(6) from the value $2.73 \times
10^{-2}$ to $ 2.43 \times 10^{-2}$. Figure 3 shows the l.h.s. of the 
modified DMO sum rule Eq.(15) (curve(a)) compared to the r.h.s. (curve (b))
after performing the above rescaling. This value of $W_{0}$ has
implications for the pion polarizability, i.e.
\begin{equation}
\alpha_{E} = \frac{\alpha_{EM}}{\mu_{\pi}} \left( \frac{<r_{\pi}^{2}>}
{3} - \frac{W_{0}}{f_{\pi}^{2}} \right) \; = 3.7 \times 10^{-4}\; 
\mbox{fm}^{3}\; ,
\end{equation}
where the numerical value above is the result of using the rescaled value
$W_{0} = 2.43 \times 10^{-2}$.
In Fig.4 we plot the l.h.s. of the 
first Weinberg sum rule Eq.(8) (curve(a)), its modified version, 
Eq.(16) (curve(b)), and their r.h.s. (curve(c)). Figure 5 shows 
the l.h.s. of the second Weinberg sum rule Eq.(9) (curve(a)), 
compared to its r.h.s. (curve(b)), and Fig. 6 the corresponding curves 
for the modified sum rule
Eq.(17). Finally, in Fig. 7 we plot the l.h.s. of the sum rule Eq.(10)
(curve(a)), the l.h.s. of the modified sum rule, Eq.(18) (curve (b)), and
their r.h.s. (curve(c)). An inspection of these results clearly indicates
that, in line with the unsatisfactory check of local duality, the original
chiral sum rules do not appear well saturated by the data. While an
overall constant rescaling of the experimental data can result in a
better saturation of the DMO sum rule, this would not help with the other
three sum rules. Hence, the problem cannot be blamed on a systematic
overall normalization uncertainty in the data. On the other hand, by
using {\it restricted global duality}, the four modified chiral sum
rules Eqs.(15)-(18) are extremely well saturated by the data.\\

In this note we have addressed the issue of the range of applicability
of perturbative QCD, augmented by condensate terms (or higher twists)
using the chiral correlator $\Pi_{V-A}$ as an example. We have argued
that the notions of duality or precocius scaling apply, in the 
space-like region, at remarkably low momentum transfers, but they must
be used with great care when time-like momenta are involved
(for discussions on the possible breakdown of local duality
in QCD see e.g. \cite{UR}-\cite{BARI}). In fact,
even at a $q^{2}$ as large as $q^{2} \simeq 3 \;\mbox{GeV}^{2}$ neither
the experimental spectral function itself, nor its moments are 
satisfactorily given by their QCD counterpart. However, by taking
linear combinations of moments, such that they vanish at the upper
limit of integration (i.e. at $s=s_{0}$, the continuum threshold),
a remarkable agreement with the QCD prediction is achieved. This we
referred to as restricted duality. Clearly, if the (finite range) 
integral over the spectral function is equated to an integral over 
a large circle in the complex energy plane, the choice of polynomial  
integral kernels that vanish at the upper limit of integration 
causes the contribution in the vicinity of the cut to 
vanish. We therefore conclude that perturbative QCD may be expected to hold
in the complex energy plane for $|q^{2}| \geq s_{0}$, with $s_{0}$ as
low as $1 \;\mbox{GeV} ^{2}$, except near the cut, where the corresponding 
radius appears to be definitely much larger than $3.5 \;\mbox{GeV} ^{2}$.\\

\begin{center}
{\bf Figure Captions}
\end{center}
Figure 1. Our fits to the vector (curve(a)) and axial-vector
(curve(b)) spectral functions  together with the  ALEPH experimental data
\cite{ALEPH}. Normalization is according to:
$\rho_{V,A} (s) = \frac{1}{2} \; \left[ 1 + {\cal{O}} \; (\alpha_{s}) 
\right] \;$ .\\
Figure 2. The left hand side of Eq.(6) computed using the fits to
the data (curve(a)), and its right hand side (curve(b))\\
Figure 3. The left hand side of Eq.(15) computed using the fits to
the data (curve(a)), and its right hand side (curve(b)) with
$- 4 \bar{L}_{10}$ rescaled to $2.43 \times 10^{- 2}$.\\
Figure 4. The left hand sides of Eq.(8) and Eq.(16) computed using
the fits to the data (curves(a)
and (b), respectively), and their right hand side (curve(c)).\\
Figure 5. The left hand side of Eq.(9) computed using the fits to
the data (curve(a)), and its right hand side (curve(b))\\
Figure 6. The left hand side of Eq.(17) computed using the fits to
the data (curve(a)), and its right hand side (curve(b))\\
Figure 7. The left hand sides of Eq.(10) and Eq.(18) computed using
the fits to the data (curves(a)
and (b), respectively), and their right hand side (curve(c)).\\
\newpage
\begin{figure}[h]
\begin{center}
\begin{picture}(400,300)(0,0)
\put(-40,360){\special{em:graph fig1.pcx}}
\end{picture}
\end{center}
\end{figure}
\newpage
\begin{figure}[h]
\begin{center}
\begin{picture}(300,200)(0,0)
\put(-95,180){\special{em:graph fig2.pcx}}
\end{picture}
\end{center}
\end{figure}
\newpage
\begin{figure}[h]
\begin{center}
\begin{picture}(300,200)(0,0)
\put(-95,180){\special{em:graph fig3.pcx}}
\end{picture}
\end{center}
\end{figure}
\newpage
\begin{figure}[h]
\begin{center}
\begin{picture}(300,200)(0,0)
\put(-95,180){\special{em:graph fig4.pcx}}
\end{picture}
\end{center}
\end{figure}
\newpage
\begin{figure}[h]
\begin{center}
\begin{picture}(300,200)(0,0)
\put(-95,180){\special{em:graph fig5.pcx}}
\end{picture}
\end{center}
\end{figure}
\newpage
\begin{figure}[h]
\begin{center}
\begin{picture}(300,200)(0,0)
\put(-95,180){\special{em:graph fig6.pcx}}
\end{picture}
\end{center}
\end{figure}
\newpage
\begin{figure}[h]
\begin{center}
\begin{picture}(300,200)(0,0)
\put(-95,180){\special{em:graph fig7.pcx}}
\end{picture}
\end{center}
\end{figure}

\begin{thebibliography}{99}
 \bibitem{CA} V. de Alfaro, S. Fubini, G. Furlan and C. Rossetti,
 Currents in Hadron Physics (North Holland, Amsterdam, 1973).
 \bibitem{PDG} Particle Data Group, R.M. Barnett et al., Phys. Rev.
 {\bf D 54} (1996) 1.
 \bibitem{GL}J. Gasser and H. Leutwyler, Nucl. Phys. {\bf B 250} (1985) 465;
 G. Ecker, J. Gasser, A. Pich, and E. de Rafael, Nucl. Phys. {\bf B 321}
 (1989) 311.
 \bibitem{AMEN} S.R. Amendolia et al., Nucl.Phys.{\bf B 277} (1986) 168.
 \bibitem{DMO} T. Das, V.S. Mathur, and S. Okubo, Phys. Rev. Lett.
 {\bf 19} (1967) 859.
 \bibitem{WSR} S. Weinberg, Phys. Rev. Lett. {\bf 18} (1967) 507.
 \bibitem{PMD} T. Das, G.S. Guralnik, V.S. Mathur, F.E. Low, and J.E. Young,
 Phys. Rev. Lett. {\bf 18} (1967) 759.
 \bibitem{RA} E.G. Floratos, S. Narison, and E. de Rafael, Nucl. Phys.
 {\bf B 155} (1979) 115.
 \bibitem{ARGUS} H. Albrecht et al., Z. Phys. {\bf C 33} (1986) 7.
 \bibitem{PS} R. Peccei and J. Sol\`{a}, Nucl. Phys. {\bf B 281} (1987) 1.
 \bibitem{CADS1} C.A. Dominguez and J. Sol\`{a}, Z. Phys. {\bf C 40}
 (1988) 63.
 \bibitem{CADS2} C.A. Dominguez and J. Sol\`{a}, Phys. Lett. {\bf B 208}
 (1988) 131.
 \bibitem{ALEPH} ALEPH Collaboration, R. Barate et al., CERN Report
 No. CERN-EP/98-12 (1998).
 \bibitem{CONOS} V. Kartvelishvili, M. Margvelashvili and G. Shaw,
 Nucl. Phys. B (Proc. Suppl.) {\bf 54 A} (1997) 309; A. H\"{o}cker,
 Orsay Report No. LAL 96-95 (1996); M. Davier, L. Girlanda, A. H\"{o}cker
 and J. Stern, Report No. HEP-PH/9802447 (1998).
 \bibitem{SVZ} M.A. Shifman, A.I. Vainshtein, V.I. Zakharov,
 Nucl. Phys. {\bf B 147} (1978) 385,448.
\bibitem{UR} M. Shifman, in Proceedings of the Workshop on Continuous
Advances in QCD, ed. A. Smilga (World Scientific, Singapore,1994) p. 249;
B. Chibisov {\it et al.} Int. J. Mod. Phys. A 12 (1997) 2075;
B. Blok, M. Shifman and Da-Xin Zhang, Phys. Rev. {\bf D 57} (1998) 2691.
\bibitem{BARI} P. Colangelo, C.A. Dominguez and G. Nardulli,
Phys. Lett. {\bf B 409} (1997) 417.
\end{thebibliography}
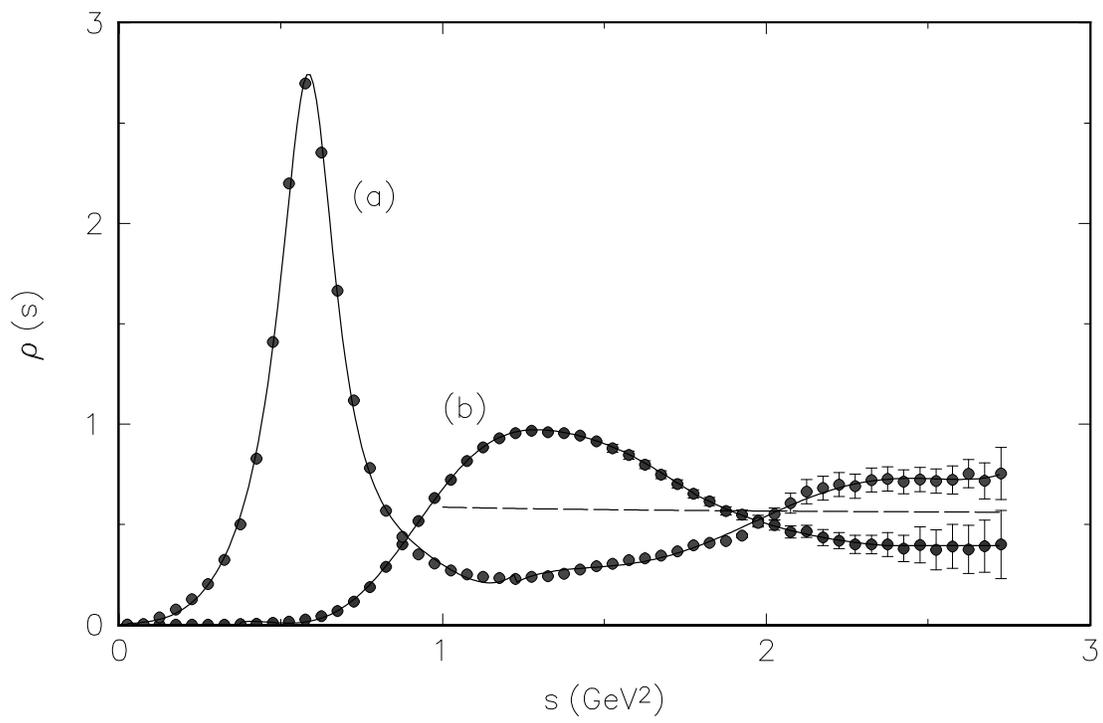
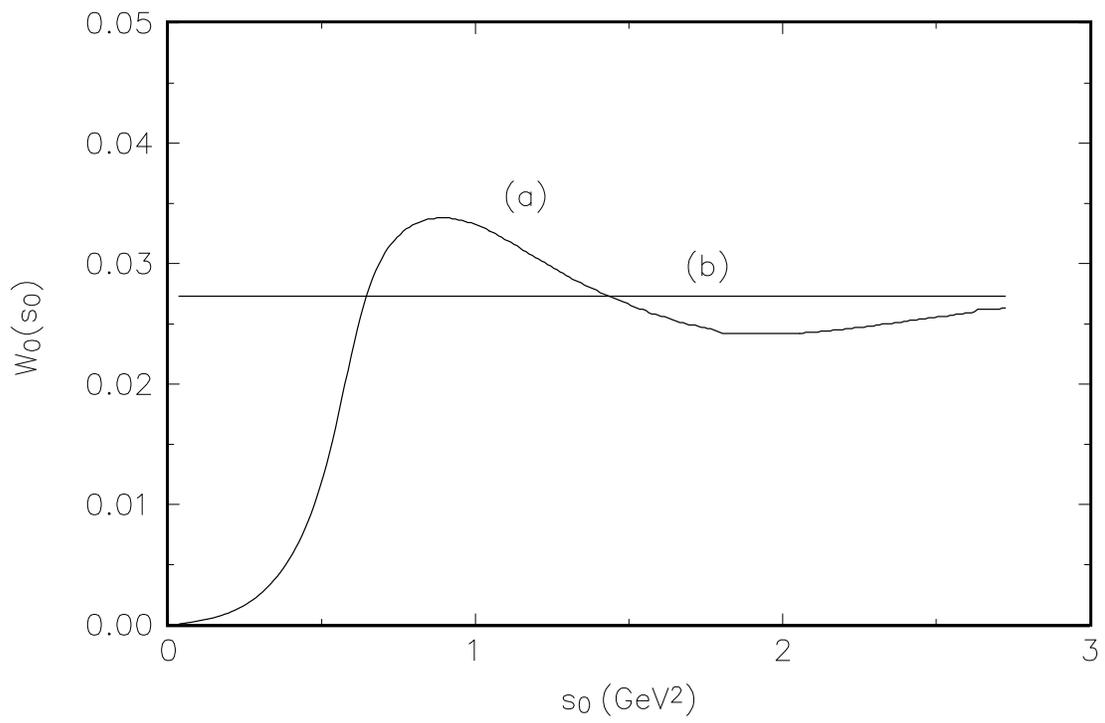
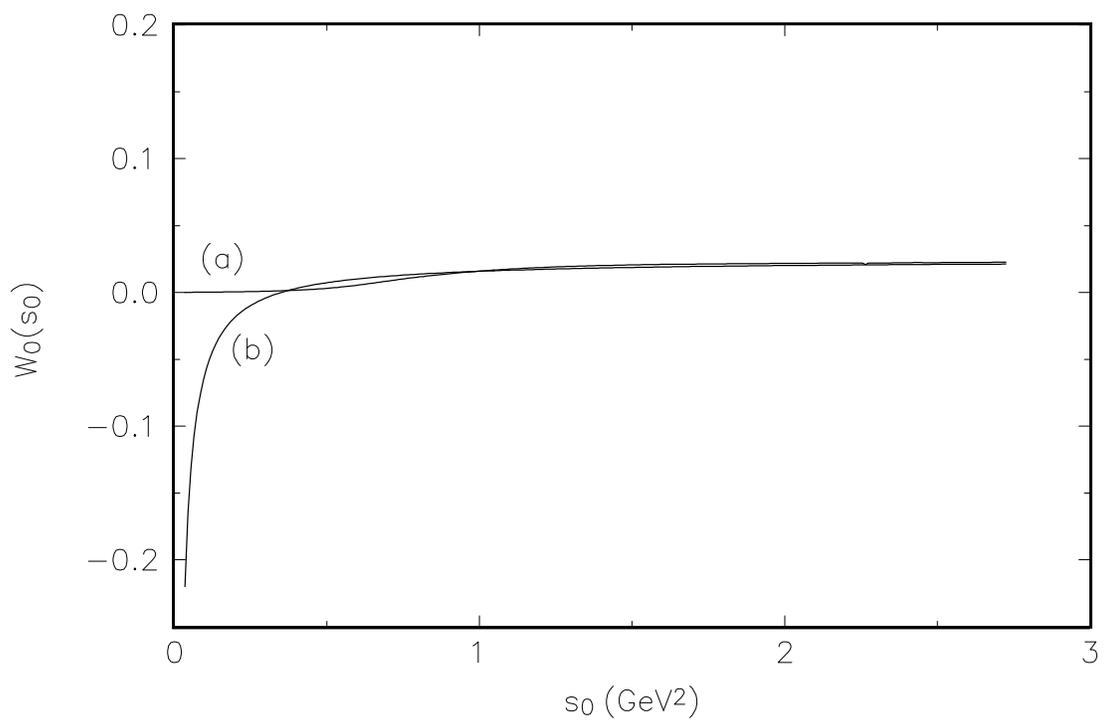
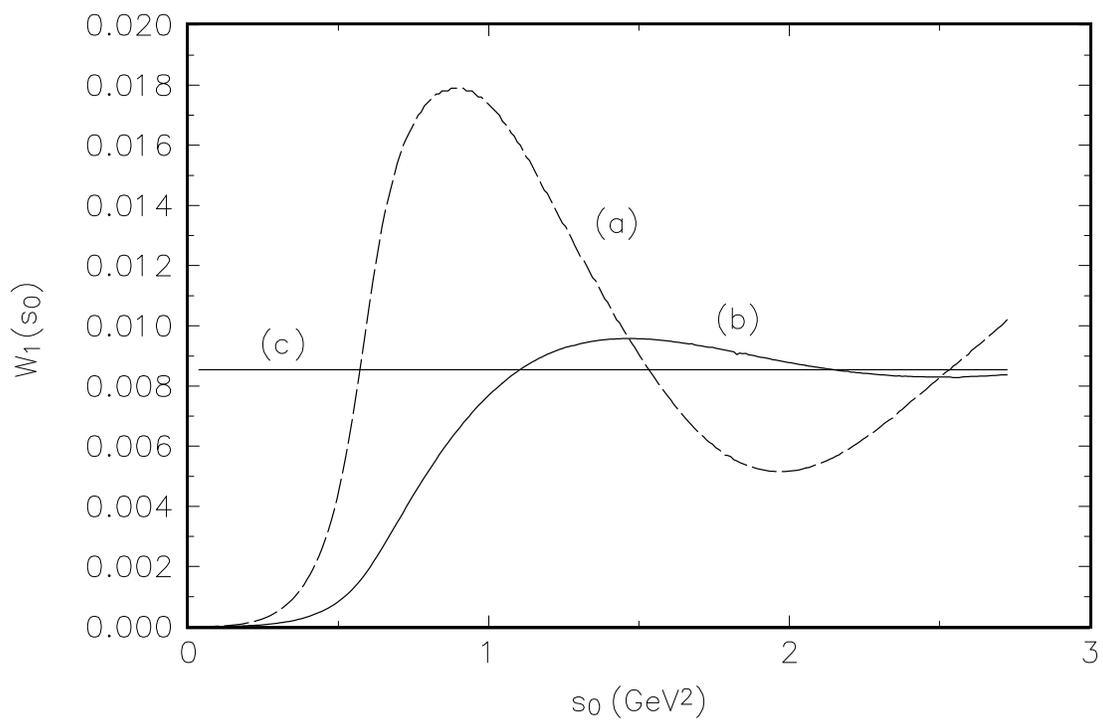
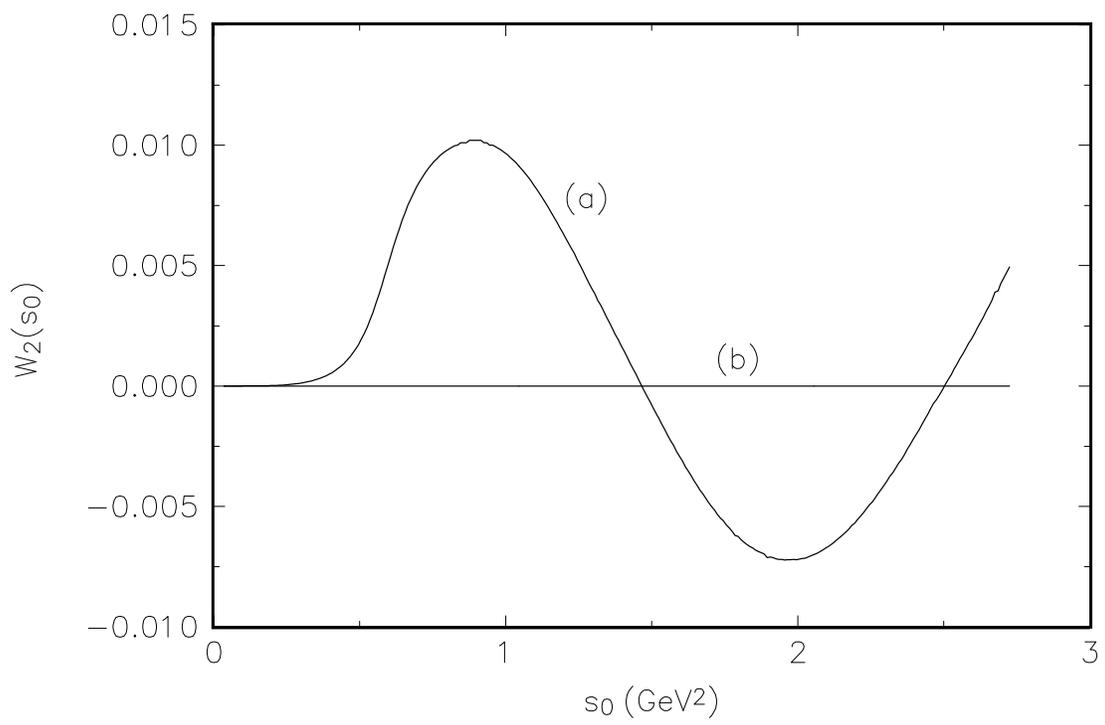
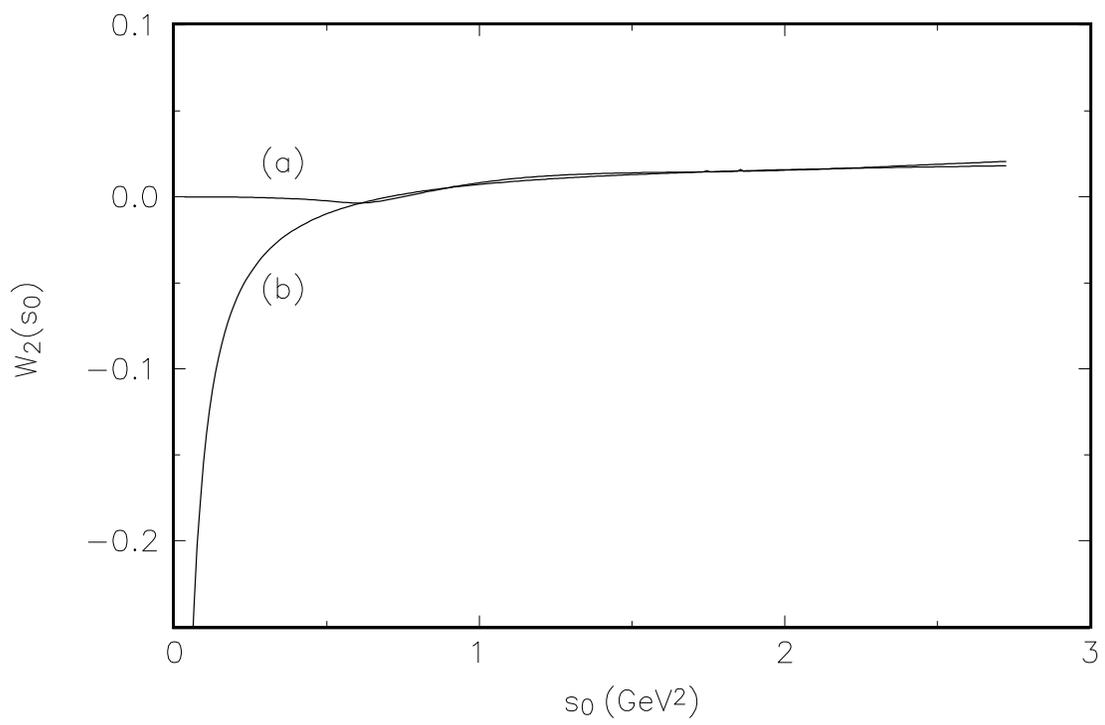
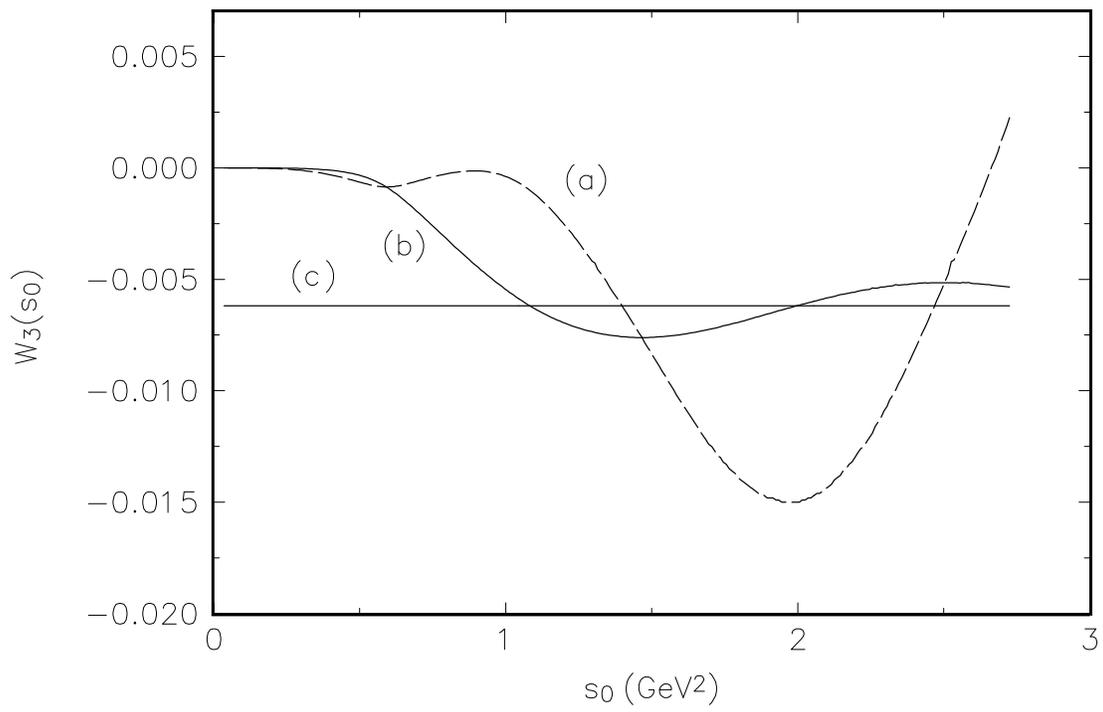
\end{document}